\documentstyle[fleqn]{article}
\begin{document}
\LARGE
\begin{center}
Cremonian Space-Time(s) as an Emergent Phenomenon
\end{center}
\vspace*{-.2cm}
\Large
\begin{center}
Metod Saniga

\vspace*{.3cm} \small {\it Astronomical Institute, Slovak Academy
of Sciences, 05960 Tatransk\' a Lomnica, Slovak Republic}
\end{center}

\vspace*{-.4cm}
\noindent
\hrulefill

\vspace*{.2cm}
\small
\noindent
{\bf Abstract}\\

\noindent 
It is shown that the notion of fundamental elements can be extended to {\it any}, i.e. not necessarily homaloidal, web of rational surfaces in a 
three-dimensional projective space. A Cremonian space-time can then be viewed as an {\it emergent} phenomenon when the condition of  ``homaloidity" 
of the corresponding web is satisfied. The point is illustrated by a couple of particular types of ``almost-homaloidal" webs of quadratic surfaces. In the first 
case, the quadrics have a line and two distinct points in common and the corresponding pseudo-Cremonian manifold is endowed with just two spatial 
dimensions. In the second case, the quadrics share six distinct points, no three of them collinear, that lie in quadruples in three different planes, and 
the corresponding pseudo-Cremonian configuration features three time dimensions. In both the cases, the limiting process of the emergence of 
generic Cremonian space-times is explicitly demonstrated.

\noindent \hrulefill

\vspace*{.5cm} \normalsize \noindent 
Although the theory of Cremonian space-time, first introduced in [1], is relatively new, it has already proven to be remarkably fertile and attracted attention
of both the lay public [2] and specialists [3] alike. The concept not only offers us a feasible explantion for why our Universe is, at the macroscopic scale, 
endowed with three spatial and one time dimensions [1,4--7], but also indicates unsuspected intricacies of a coupling between the two [6]. Moreover, it 
gives us important qualitative hints as for a possible underlying algebraic geometrical structure of a large variety of non-ordinary forms of psychological 
time (and space) [6,8]. These fascinating properties alone are enough to realize that the theory deserves further serious exploration.

One of the most natural and fruitful ways of getting a deeper insight into a(ny) theory is to relax one (or several) assumptions that the theory 
is based on and see what structural and conceptual changes such a step involves. In order to pursue this strategy in our case one has to recall
the underlying geometrical principle behind our concept of Cremonian space-time: the existence of a {\it homaloidal} web of (quadric) surfaces in
a three-dimensonial projective space, $P_{3}$ [1,4,5]. An aggregate of surfaces (of any order, not necessarily quadratic) in $P_{3}$ is homaloidal 
[1,9]:  if a) it is linear and of freedom three (i.e. contains a triple infinity of surfaces), b) all its surfaces are rational, and c) any 
three distinct members of the set have only one free (variable) intersection. The generalized theory outlined in what follows is based on abandoning
the last assumption, i.e. on allowing any three distinct surfaces of the web to have two (or more) points in common.

Of the couple of different types of homaloidal web of quadrics we
have so far dealt with we shall first consider that whose corresponding Cremonian space-time was found to mimic best the basic observed macroscopic
properties of the Universe, viz. a web whose base configuration comprises a (straight-)line and three distinct, non-collinear points, none incident
with the line in question [1,6]. So, in this case, any web of quadrics that features a base line and where the number of isolated base points falls short
of three may serve our purpose. 

We shall, naturally, focus on the case where there are just two base points as this case represents, obviously, the smallest possible deviation
from the ``homaloidity" condition. In order to facilitate our reasoning, we shall choose a system of homogeneous coordinates $\breve{z}_{i}$,
$i=1,2,3,4,$ in which the equation of base line, $\widehat{{\cal L}}$, reads
\begin{equation}
\widehat{{\cal L}}:~~\breve{z}_{1} = 0 = \breve{z}_{2}
\end{equation}
and the two isolated base points, $\widehat{\rm B}_{1}$ and $\widehat{\rm B}_{2}$, coincide, respectively, with the vertices V$_{1}$ and V$_{2}$ of the
coordinate tetrahedron, i.e.
\begin{equation}
\widehat{\rm B}_{1}:~~\varrho \breve{z}_{i} = (1, 0, 0, 0),
\end{equation}
\begin{equation}
\widehat{\rm B}_{2}:~~\varrho \breve{z}_{i} = (0, 1, 0, 0),
\end{equation}
where $\varrho$ is, in what follows, a non-zero proportionality factor. Now, employing the equation of a generic quadric, ${\cal D}$, of $P_{3}$,
\begin{eqnarray}
{\cal D} \equiv \sum_{i,j=1}^{4} d_{ij} \breve{z}_{i} \breve{z}_{j} = 
d_{11} \breve{z}_{1}^{2} + d_{22} \breve{z}_{2}^{2} + d_{33} \breve{z}_{3}^{2} + d_{44} \breve{z}_{4}^{2} +
2d_{12} \breve{z}_{1}\breve{z}_{2} + 2d_{13} \breve{z}_{1}\breve{z}_{3} + 2d_{14} \breve{z}_{1}\breve{z}_{4} + \hspace*{0.35cm} \nonumber \\ 
+~ 2d_{23} \breve{z}_{2}\breve{z}_{3} + 2d_{24} \breve{z}_{2}\breve{z}_{4} + 2d_{34} \breve{z}_{3}\breve{z}_{4} = 0,
\end{eqnarray}
we find that the system of quadrics that contain $\widehat{{\cal L}}$ and pass through $\widehat{\rm B}_{1}$ and $\widehat{\rm B}_{2}$ is given by
\begin{equation}
d_{12} \breve{z}_{1}\breve{z}_{2} + d_{13} \breve{z}_{1}\breve{z}_{3} + d_{14} \breve{z}_{1}\breve{z}_{4} 
+ d_{23} \breve{z}_{2}\breve{z}_{3} + d_{24} \breve{z}_{2}\breve{z}_{4} = 0,
\end{equation}
where each $d$'s may acquire any real value. This aggregate is, however, not a web for it effectively  depends on four, instead of
three, parameters. In order to get a web from it, a linear constraint has to be imposed on the parameters $d$'s. As $d_{12}\neq 0$ --
otherwise the aggregate would contain another base line (the $\breve{z}_{3} = 0 = \breve{z}_{4}$ one) -- this constraint can be written,
without any loss of generality, in the form
\begin{equation}
d_{12} =  \kappa_{1} d_{13}  + \kappa_{2} d_{23} + \kappa_{3} d_{14} + \kappa_{4} d_{24},
\end{equation}
where $\kappa_{i}$, $i=1,2,3,4$,  are regarded as fixed constants. After substituting the last equation into Eq.\,(5), we get the web desired 
\begin{eqnarray}
{\cal W}^{\clubsuit}(\vartheta) &=& \sum_{i=1}^{4} \vartheta_{i}{\cal D}_{i}^{\clubsuit} \nonumber \\
&\equiv& \vartheta_{1} \breve{z}_{1} (\breve{z}_{3} + \kappa_{1} \breve{z}_{2})
+ \vartheta_{2} \breve{z}_{2} (\breve{z}_{3} + \kappa_{2} \breve{z}_{1})
+ \vartheta_{3} \breve{z}_{1} (\breve{z}_{4} + \kappa_{3} \breve{z}_{2})
+\vartheta_{4} \breve{z}_{2} (\breve{z}_{4} + \kappa_{4} \breve{z}_{1}) = 0,
\end{eqnarray}
where we simplified the notation by putting $d_{13} \equiv \vartheta_{1}$, $d_{23} \equiv \vartheta_{2}$, $d_{14} \equiv \vartheta_{3}$ and 
$d_{24} \equiv \vartheta_{4}$.

At this point we recall the definition of a Cremonian space-time as a configuration composed of the totality of {\it fundamental} elements associated with 
a homaloidal web [1,4--7]. A fundamental element of a given homaloidal web is any algebraic geometrical object (a curve, or a surface) whose only 
intersections with a member of the web are the base elements of the latter [9]. In the case of quadrics, the fundamental elements are of two distinct kinds, 
namely lines and conics, and form pencils, i.e. linear, singly-parametrical aggregates/systems. A pencil of lines is taken to generate/represent a 
macroscopic dimension of space, while that of conics -- time. From its definition it readily follows that the concept of a fundamental
element, and so that of a Cremonian space-time, is not tied solely to homaloidal webs, but it can be extended perfectly to {\it any} web 
whatsoever! A crucial property of fundamental elements is that they are located on the Jacobian surface of the web [9], i.e. on the surface
formed by the totality of vertices of the cones contained in the web; or, what amounts to the same, the totality of singular points of 
degenerate/composite quadrics in the web. 

Our forthcoming task is thus to find the form of the Jacobian surface, ${\cal J}$, for the web given by Eq.\,(7). From the above-given properties it
follows that given a generic web of quadrics,
\begin{equation}
{\cal W}(\vartheta) = \sum_{i=1}^{4} \vartheta_{i}{\cal D}_{i} = 0 ,
\end{equation}
its  ${\cal J}$ is the locus of points satisfying the following equation [9]
\begin{equation}
{\cal J} = \det \left(
\begin{array}{cccc}
\partial {\cal D}_{1}/\partial \breve{z}_{1} &  \partial {\cal D}_{2}/\partial \breve{z}_{1}& 
\partial {\cal D}_{3}/\partial \breve{z}_{1} & \partial {\cal D}_{4}/\partial \breve{z}_{1} \\
\partial {\cal D}_{1}/\partial \breve{z}_{2} &  \partial {\cal D}_{2}/\partial \breve{z}_{2}& 
\partial {\cal D}_{3}/\partial \breve{z}_{2} & \partial {\cal D}_{4}/\partial \breve{z}_{2} \\
\partial {\cal D}_{1}/\partial \breve{z}_{3} &  \partial {\cal D}_{2}/\partial \breve{z}_{3}& 
\partial {\cal D}_{3}/\partial \breve{z}_{3} & \partial {\cal D}_{4}/\partial \breve{z}_{3} \\
\partial {\cal D}_{1}/\partial \breve{z}_{4} &  \partial {\cal D}_{2}/\partial \breve{z}_{4}& 
\partial {\cal D}_{3}/\partial \breve{z}_{4} & \partial {\cal D}_{4}/\partial \breve{z}_{4} 
\end{array}
\right) = 0,
\end{equation}
which when combined with Eq. (7) yields
\begin{equation}
{\cal J}^{\clubsuit} = 2 \breve{z}_{1} \breve{z}_{2} \overline{{\cal D}}^{\clubsuit} = 0,
\end{equation}
where
\begin{equation}
\overline{{\cal D}}^{\clubsuit} \equiv \breve{z}_{4}(\kappa_{1} \breve{z}_{2} - \kappa_{2} \breve{z}_{1}) -
\breve{z}_{3}(\kappa_{3} \breve{z}_{2} - \kappa_{4} \breve{z}_{1}).
\end{equation}
It is a quite straightforward task for the reader to verify that the fundamental elements of web (7) are, in a general case, located only in the
planes $\breve{z}_{2}$ and $\breve{z}_{1}$; in both the cases these elements are lines forming pencils centered, respectively, at
$\widehat{\rm B}_{1}$, viz.
\begin{equation}
\widetilde{\cal L}_{1}(\vartheta):~~\breve{z}_{2} = 0 = \vartheta_{1} \breve{z}_{3} + \vartheta_{3}\breve{z}_{4}, 
\end{equation}
and $\widehat{\rm B}_{2}$, viz.
\begin{equation}
\widetilde{\cal L}_{2}(\vartheta):~~\breve{z}_{1} = 0 = \vartheta_{2} \breve{z}_{3} + \vartheta_{4}\breve{z}_{4};
\end{equation}
for a line meets a quadric in two (not necessarily distinct and/or real) points and these are in our cases furnished, respectively, by
$\widehat{\rm B}_{1}$ and $\widehat{\rm B}_{2}$ and a point of $\widehat{{\cal L}}$. The quadric $\overline{{\cal D}}^{\clubsuit} = 0$
contains, in general, only one further fundamental element apart from a couple of lines shared with $\widetilde{\cal L}_{1}(\vartheta)$ and
$\widetilde{\cal L}_{2}(\vartheta)$, namely the line joining the points  $\widehat{\rm B}_{1}$ and $\widehat{\rm B}_{2}$. Yet, this quadric is
the most interesting piece of ${\cal J}^{\clubsuit}$ because when it becomes {\it composite} (singular), the web ${\cal W}^{\clubsuit}$ becomes 
{\it homaloidal} and the quadric itself exhibits two different, singly-infinite aggregates of fundamentals.

In order to see that explicitly, one recalls [1,9] that a quadric ${\cal D}$, Eq.\,(4), is composite iff
\begin{equation}
\det \left(
\begin{array}{cccc}
d_{11} & d_{12} & d_{13} & d_{14} \\
d_{21} & d_{22} & d_{23} & d_{24} \\
d_{31} & d_{32} & d_{33} & d_{34} \\
d_{41} & d_{42} & d_{43} & d_{44} 
\end{array}
\right) = 0
\end{equation}
and, for Eq.\,(11), this equation reduces to
\begin{equation}
\kappa_{1} \kappa_{4} - \kappa_{2} \kappa_{3} = 0. 
\end{equation}   
To find the form of this composite quadric, $\overline{{\cal D}}^{\clubsuit}_{\odot}$, we may assume, without any substantial loss of generality,
that $\kappa_{1}$ is non-zero, rewrite Eq.\,(15) as
\begin{equation}
 \kappa_{4} = \frac{\kappa_{2} \kappa_{3}}{\kappa_{1}} 
\end{equation}
and insert the latter into Eq.\,(11) to arrive at
\begin{equation}
\overline{{\cal D}}^{\clubsuit}_{\odot} = \frac{1}{\kappa_{1}} (\kappa_{1} \breve{z}_{2} - \kappa_{2} \breve{z}_{1})
( \kappa_{1} \breve{z}_{4} - \kappa_{3} \breve{z}_{3});
\end{equation}
that is, $\overline{{\cal D}}^{\clubsuit}_{\odot}$ consists of a pair of {\it planes}, $\kappa_{1} \breve{z}_{2} - \kappa_{2} \breve{z}_{1} = 0$ and
$\kappa_{1} \breve{z}_{4} - \kappa_{3} \breve{z}_{3} = 0$. As Eq.\,(7) acquires under Eq.\,(16) the form
\begin{eqnarray}
{\cal W}^{\clubsuit}_{\odot}(\vartheta) &=& \vartheta_{1} \breve{z}_{1} (\breve{z}_{3} + \kappa_{1} \breve{z}_{2})
+ \vartheta_{2} \breve{z}_{2} (\breve{z}_{3} + \kappa_{2} \breve{z}_{1})
+ \vartheta_{3} \breve{z}_{1} (\breve{z}_{4} + \kappa_{3} \breve{z}_{2})
+\vartheta_{4} \breve{z}_{2} \left(\breve{z}_{4} + \frac{\kappa_{2} \kappa_{3}}{\kappa_{1}} \breve{z}_{1}\right) \nonumber \\
&=& 0,
\end{eqnarray}
it is easy to spot that this ``constrained" web contains, indeed, a {\it third} base point, viz.
\begin{equation}
\widehat{\rm B}_{3}^{\odot}:~~\varrho \breve{z}_{i} = ( \kappa_{1}, \kappa_{2}, - \kappa_{1} \kappa_{2}, - \kappa_{2}\kappa_{3}),
\end{equation}
and is thus homaloidal [1,6,9], featuring, in addition to $\widetilde{\cal L}_{1}(\vartheta)$ and
$\widetilde{\cal L}_{2}(\vartheta)$, one more pencil of fundamental {\it lines}, viz.
\begin{equation}
\widetilde{\cal L}_{3}^{\odot}(\vartheta):~~\kappa_{1} \breve{z}_{2} - \kappa_{2} \breve{z}_{1} = 0 = \left(\frac{\kappa_{1}}{\kappa_{2}} \vartheta_{1} 
+ \vartheta_{2}\right)(\breve{z}_{3} + \kappa_{1} \breve{z}_{2}) +
\left(\frac{\kappa_{1}}{\kappa_{2}} \vartheta_{3} 
+ \vartheta_{4}\right)(\breve{z}_{4} + \kappa_{3}\breve{z}_{2}),
\end{equation}
and a pencil of fundamental {\it conics}, viz.
\begin{equation}
\widetilde{\cal Q}^{\odot}(\vartheta):~~\kappa_{1} \breve{z}_{4} - \kappa_{3} \breve{z}_{3} = 0 = 
\left( \vartheta_{1} + \frac{\kappa_{3}}{\kappa_{1}}\vartheta_{3}\right)\breve{z}_{1} (\breve{z}_{3} + \kappa_{1} \breve{z}_{2}) +
\left(\vartheta_{2} + \frac{\kappa_{3}}{\kappa_{1}}\vartheta_{4}\right)\breve{z}_{2}(\breve{z}_{3} + \kappa_{2} \breve{z}_{1});
\end{equation}
the two aggregates being located, as expected, in the two sheets of $\overline{{\cal D}}^{\clubsuit}_{\odot}$.
Our findings can be rephrased as follows. The ``pseudo-", or ``proto-"Cremonian configuration associated with a generic   
${\cal W}^{\clubsuit}$ consists of two space dimensions ($\widetilde{\cal L}_{1}(\vartheta)$ and
$\widetilde{\cal L}_{2}(\vartheta)$) and transforms into a fully-developed ``classical" Cremonian space-time [1,3,6], endowed with an 
additional space dimension ($\widetilde{\cal L}_{3}^{\odot}(\vartheta)$) and time ($\widetilde{\cal Q}^{\odot}(\vartheta)$), whenever the web 
becomes homaloidal. This phenomenon can be given another nice geometrical picture. We may regard $\kappa_{i}$'s as 
homogenous coordinates of a variable point in an abstract three-dimensional projective space. Eq.\,(15) (or, equivalently, Eq.\,(16)) then 
defines a quadric surface in this space and the ``homaloidity" condition simply answers to the fact that the point happens to fall on this quadric. 

In order to get a deeper insight into the nature of this ``emergence phenomenon", we shall consider a web of quadrics through the following
six distinct points:
\begin{eqnarray}
\widehat{\rm B}_{1}:~~\varrho \breve{z}_{i} = (1, 0, 0, 0),~~~~~\widehat{\rm B}_{4}:~~\varrho \breve{z}_{i} = (\kappa, 0, 0, 1), \nonumber \\
\widehat{\rm B}_{2}:~~\varrho \breve{z}_{i} = (0, 1, 0, 0),~~~~~\widehat{\rm B}_{5}:~~\varrho \breve{z}_{i} = (0, \kappa, 0, 1),  \\
\widehat{\rm B}_{3}:~~\varrho \breve{z}_{i} = (0, 0, 1, 0),~~~~~\widehat{\rm B}_{6}:~~\varrho \breve{z}_{i} = (0, 0, \kappa, 1), \nonumber
\end{eqnarray}
where $\kappa$ is a variable real parameter. These points are, obviously, all real, lying in quadruples in three diferent planes and no three on
the same line.  The web they define is of the form
\begin{equation}
{\cal W}^{\spadesuit}(\vartheta) = \vartheta_{1} \breve{z}_{2}\breve{z}_{3} + \vartheta_{2} \breve{z}_{1}\breve{z}_{3} + 
\vartheta_{3} \breve{z}_{1}\breve{z}_{2} 
+ \vartheta_{4} \breve{z}_{4}(\breve{z}_{1}+\breve{z}_{2}+\breve{z}_{3} - \kappa \breve{z}_{4}) = 0
\end{equation}
and its Jacobian reads
\begin{equation}
{\cal J}^{\spadesuit} = 2 \breve{z}_{1} \breve{z}_{2} \breve{z}_{3}(\breve{z}_{1}+\breve{z}_{2}+\breve{z}_{3} - 2\kappa \breve{z}_{4}) = 0.
\end{equation}
Although the Jacobian features four distinct planes, in a general case, $\kappa \neq 0$, only three of them carry sets of fundamental elements. 
These are
the planes $\breve{z}_{1} = 0$, $\breve{z}_{2}=0$ and $\breve{z}_{3}=0$, and the corresponding fundamental elements are {\it 
conics} in the following pencils:
\begin{equation}
\widetilde{{\cal Q}}_{1}(\vartheta):~~\breve{z}_{1} = 0 = \vartheta_{1}\breve{z}_{2}\breve{z}_{3}
+ \vartheta_{4} \breve{z}_{4}(\breve{z}_{2}+\breve{z}_{3} - \kappa \breve{z}_{4}),
\end{equation}
\begin{equation}
 \widetilde{{\cal Q}}_{2}(\vartheta):~~\breve{z}_{2} = 0 = \vartheta_{2}\breve{z}_{1}\breve{z}_{3}  
+ \vartheta_{4} \breve{z}_{4}(\breve{z}_{1}+\breve{z}_{3} - \kappa \breve{z}_{4}),
\end{equation}
and
\begin{equation}
 \widetilde{{\cal Q}}_{3}(\vartheta):~~\breve{z}_{3} = 0 = \vartheta_{3}\breve{z}_{1}\breve{z}_{2} 
+ \vartheta_{4} \breve{z}_{4}(\breve{z}_{1}+\breve{z}_{2} - \kappa \breve{z}_{4}),
\end{equation}
respectively. The conics (of a triply-infinite system),
\begin{equation}
 \widetilde{{\cal Q}}^{\star}(\vartheta):~~\breve{z}_{1}+\breve{z}_{2}+\breve{z}_{3} - 2\kappa \breve{z}_{4}= 0 =  \vartheta_{1}\breve{z}_{2}\breve{z}_{3} +
\vartheta_{2}\breve{z}_{1}\breve{z}_{3} 
 + \vartheta_{3}\breve{z}_{1}\breve{z}_{2}
+ \kappa \vartheta_{4} \breve{z}_{4}^{2},
\end{equation}
cut out from ${\cal W}^{\spadesuit}$ by the remaining Jacobian plane, $\breve{z}_{1}+\breve{z}_{2}+\breve{z}_{3} - 2\kappa \breve{z}_{4}=0$, 
cannot be fundamental elements as they do not contain any of the base points (see Eq.\,(22)). ${\cal W}^{\spadesuit}$ becomes homaloidal
for $\kappa \rightarrow 0$, 
\begin{equation}
{\cal W}^{\spadesuit}_{\odot}(\vartheta) \equiv {\cal W}^{\spadesuit}_{\kappa \rightarrow 0}(\vartheta)  =  \vartheta_{1}\breve{z}_{2}\breve{z}_{3} + 
\vartheta_{2}\breve{z}_{1}\breve{z}_{3} + 
 \vartheta_{3}\breve{z}_{1}\breve{z}_{2}
+ \vartheta_{4} \breve{z}_{4}(\breve{z}_{1}+\breve{z}_{2}+\breve{z}_{3}) = 0,
\end{equation}
in which case, in addition to the three pencils of fundamental conics
\begin{equation}
\widetilde{{\cal Q}}_{\alpha}^{\odot}(\vartheta) \equiv \widetilde{{\cal Q}}_{\alpha}^{\kappa \rightarrow 0}(\vartheta):~~\breve{z}_{\alpha} = 0 = 
\vartheta_{\alpha}\breve{z}_{\beta}\breve{z}_{\gamma}
+ \vartheta_{4} \breve{z}_{4}(\breve{z}_{\beta}+\breve{z}_{\gamma}),~~\alpha \neq \beta \neq \gamma,~~\alpha, \beta, \gamma =1,2,3,
\end{equation}
we also have a pencil of fundamental {\it lines}, namely
\begin{equation}
\widetilde{{\cal L}}^{\odot}(\vartheta) \equiv \widetilde{{\cal Q}}^{\star}_{\kappa \rightarrow 0}(\vartheta):~~\breve{z}_{1}+\breve{z}_{2}+\breve{z}_{3} = 0 
= \vartheta_{1}\breve{z}_{2}\breve{z}_{3} +
\vartheta_{2}\breve{z}_{1}\breve{z}_{3} 
+ \vartheta_{3}\breve{z}_{1}\breve{z}_{2};
\end{equation}
these lines share point V$_{4}$ ($\varrho \breve{z}_{i} = (0,0,0,1)$), the common merger of three of the base points, viz. $\widehat{\rm B}_{4}$,
$\widehat{\rm B}_{5}$ and $\widehat{\rm B}_{6}$, and the point at which all the quadrics of  ${\cal W}^{\spadesuit}_{\odot}$ touch the plane 
$\breve{z}_{1}+\breve{z}_{2}+\breve{z}_{3} = 0$ [5,9]. Although this emerging Cremonian space-time features three time dimensions
($\widetilde{{\cal Q}}_{\alpha}^{\odot}(\vartheta)$) and a single spatial one ($\widetilde{{\cal L}}^{\odot}(\vartheta)$), and is thus an inverse to that
offered to our senses, the case itself serves as an important illustration of the intricacy of the coupling between the {\it extrinsic} structure of time 
(i.e. the {\it type} of a pencil of conics) and the {\it number} of spatial coordinates. For the three time dimensions of our pseudo-Cremonian 
manifold, $\widetilde{{\cal Q}}_{\alpha}(\vartheta)$, represent each a generic pencil of conics, i.e. a pencil endowed with four distinct base points, 
while those of the limiting Cremonian sibling, $\widetilde{{\cal Q}}_{\alpha}^{\odot}(\vartheta)$, are each a pencil with three distinct base points only, 
namely $\widehat{\rm B}_{\beta}$, $\widehat{\rm B}_{\gamma}$ and V$_{4}$, the last one being of multiplicity two [see Figs.\,$1a,b$ of Ref.\,6].  
One thus sees that the ``birth" of a space dimension, $\widetilde{{\cal L}}^{\odot}(\vartheta)$, entails serious structural changes 
in all the three time coordinates. This feature dovetails nicely with what we found for strictly homaloidal transitions [6], characterized, however, 
by a drop in the number of spatial dimensions. 

From this reasoning it is obvious that a Cremonian space-time is a rather exceptional structure, whose emergence is of a fairly
complex nature. Following the strategy employed above, it should represent no difficulty for the interested reader to examine other
potential transitions,
in particular those where pseudo-Cremonian configurations feature only a finite (or even zero) number of elements. Insights might also be obtained from 
an analysis of (pseudo-)Cremonian space-times associated with webs of cubic and/or higher order surfaces.
All this implies a wealth of additional possibilities to those outlined in [7] regarding ``Cremonian" scenarios of how our Universe might have come into
being. 
Here, we are confronted with a fascinating possibility that the Universe may have spent a substantial fraction of its life-time in some pseudo-Cremonian 
regime and acquired its current generic ``quadro-cubic" Cremonian form [1,6,7] only ``relatively recently". This intriguing 
scenario will be examined in more detail in a separate paper. 
\\ \\ 
{\bf Acknowledgements}\\ 
I am grateful to Mark Stuckey (Elizabethtown College, PA) for a proofreading of the paper. The work was supported in part by a
2001--2002 NATO/FNRS Advanced Research Fellowship and the 2001--2003 CNR-SAV joint research project ``The Subjective Time
and its Underlying Mathematical Structure."

\end{document}